\documentclass[prl,twocolumn,letterpaper,showpacs,superscriptaddress]{revtex4}
\def\be{\begin{equation}}
\def\fin{\end{equation}}
\def\disp{\displaystyle}

\def\T{{\sf T\kern-.45em T}}
\def\C{\kern.1em{\raise.47ex\hbox{$\scriptscriptstyle |$}}
             \kern-.40em{\sf C}}

\def\hfl{\disp\mathop{\hbox to 10mm{\rightarrowfill}}}

\usepackage{graphicx,amssymb,amsmath}
\begin{document}
\title{Optimal search strategies for hidden targets}

\date{\today}

\author{O. B\'enichou}
\affiliation{Laboratoire de Physique Th{\'e}orique des Liquides,
Universit{\'e} Paris 6, 4 Place Jussieu, 75252 Paris, France}
\author{M.Coppey}
\affiliation{Laboratoire de Physique Th{\'e}orique des Liquides,
Universit{\'e} Paris 6, 4 Place Jussieu, 75252 Paris, France}
\author{M.Moreau}
\affiliation{Laboratoire de Physique Th{\'e}orique des Liquides,
Universit{\'e} Paris 6, 4 Place Jussieu, 75252 Paris, France}
\author{P-H. Suet}
\affiliation{Laboratoire de Physique Th{\'e}orique des Liquides,
Universit{\'e} Paris 6, 4 Place Jussieu, 75252 Paris, France}\author{R.Voituriez}
\affiliation{Physicochimie Curie (CNRS-UMR168), Institut Curie, Section 
de Recherche, 
26 rue d'Ulm 75248 Paris Cedex 05 France}

\begin{abstract}What is the fastest way of finding a randomly hidden target? This question of general relevance is of vital importance for foraging animals. Experimental observations reveal that the search behaviour of foragers is generally  intermittent: active search phases randomly alternate with phases of fast ballistic motion. In this letter, we study the efficiency of this  type of two states  search strategies, by calculating analytically the mean first passage time at the target. We  model the perception mecanism involved in the active search phase  by a diffusive process. In this framework,  we show that the search strategy is optimal when the average duration of "motion phases" varies like the power either 3/5 or 2/3 of the average duration of "search phases", depending on the regime. This scaling accounts for experimental data over a wide range of species, which suggests that the kinetics of search trajectories is a determining factor optimized by foragers and that the perception activity is adequately described by a diffusion process. 
\end{abstract}


\maketitle

 Searching for a randomly located object is one of the most frequent  tasks of living organisms, be it    for obtaining food, a sexual partner or a shelter\cite{r1}. In these examples, the search time is generally a limiting factor which has to be optimized for the  survival of the species. The question of determining the efficiency of  a search behaviour    is thus  a crucial problem of behavioral ecology, which has inspired numerous experimental\cite{r1,r2,r3,r4,r5} and theoretical\cite{r7,r8,Klafter,krebs,r6} works . It is also relevant to broader domains such as stochastic processes theory\cite{Stan1,Stan2}, applied mathematics\cite{math} and molecular biology\cite{r18,r15}.  
 
 Anyone who has ever lost his keys knows that instinctively we adopt an intermittent behaviour  combining local scanning phases and relocating phases.
Indeed, numerous studies of foraging behaviour of a broad range of animal species show that such an  intermittent behaviour is commonly observed and that the durations of search and displacement phases vary widely\cite{r1,r2,r3}. The spectrum, which goes from cruise strategy (ex. for large fishes that swim continuously such as tuna), to ambush or sit-and-wait search, where the forager remains stationary for long periods (such as rattlesnake), has never been interpreted quantitatively. The intermittent strategy, often referred to as "saltatory"\cite{r2,r3}, can be understood intuitively when the targets are "difficult" to detect and sparsely distributed, as it is the case for many
foragers (such as ground foraging birds, lizards, planktivorous fish...): since a fast movement is known to significantly degrade perception abilities\cite{r2,r3}, the forager must search slowly. Then, it has to relocate as fast as possible in order to explore a previously unscanned space, and search slowly again. 

Even though numerous models based on optimization of the net energy gain\cite{r4,r5,r6} predict an optimal strategy for foragers, the large number of unknown parameters used to model the complexity of the energetic constraint, renders a quantitative comparison with experimental data difficult. Here, as has already been suggested\cite{r7,r8}, we assume that the search time is the relevant quantity optimized by the forager in order to obtain a sufficient daily amount of food and to precede other competing foragers. We treat the energy cost only as an external constraint that sets the maximal speed of the animal. We develop a general purely kinetic model of target search, which captures the essential features of saltatory search behaviour observed for foragers in experiments\cite{r2}, when the predator has no information about the prey location.

\begin{figure}

\scalebox{2.7}{
\includegraphics{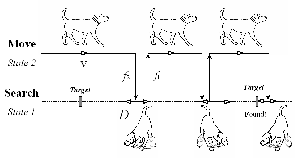}}
\caption{A two states model of saltatory search behaviour.}
\label{fig1}
\end{figure}

The central point of our schematic model (see figure \ref{fig1}) is that it relies on the explicit description of searching trajectories. In particular, as we will show,  it  permits us to elucidate the nature of the search phase. In the following we assume that the searcher displays alternatively two distinct attitudes:

(1) a search phase, hereafter referred to as phase 1, during which the searcher explores its immediate vicinity using its sensory organs. As justified below, this local scanning is modelled as a "slow" diffusive movement (a continuous random walk with diffusion coefficient $D$).  The target is found when this movement reaches the target location for the first time. 

(2)  a motion phase, referred to as phase 2, during which the searcher moves "fast" and is unable to detect a target. These repositioning moves are characterized by a ballistic motion (at constant velocity $v$). 

Next, We  assume that the searcher randomly switches from phase 1 (resp. 2) to phase 2 (resp. 1) with a rate per unit time $f_1$ (resp. $f_2$), and that the preys are immobile and randomly distributed with uniform density. Note that the average durations of phases 1 and 2 are then $1/f_1$  and $1/f_2$. The stochastic nature of the model is justified in the case of foragers by the observation of large fluctuations of the duration of search and motion phases for each animal\cite{r3}. Indeed,  exponential laws are widely observed for animal behaviours\cite{four}. Our goal is to study the optimal search strategy with respect to its kinetic mechanism only, i.e. to determine the rates $f_1$, $f_2$  which minimize the first passage time of the searcher at a target location.

Let us justify briefly our modelling of each phase. It is clearly beyond the scope of our model to describe the sensory search phase in details, as it involves very complex biological mechanisms. Our goal here is to capture the main features of this activity which are relevant from a kinetic point of view.  For hidden targets, the stimuli emitted are very weak and the predator is very likely not to detect them. Then it may have to scan many times the same location before finding the target: we model this mechanism by a diffusion process. Note that in this description, the successive positions of the diffusive trajectory are not necessarily the very positions of the animal, but the focus points of the sense involved. This idea of a diffusion process (or more generally of random walk process) for sensory perception has already been suggested, in particular for vision\cite{r9}, tactile sense or olfaction\cite{r1}. As focusing and processing the information received by sensory organs require a minimum time, the search phase can not be to short, which implies an upper bound  $f_{1{\rm max}}$   for the rate $f_1$. Since the objective of phase 2 moves is to explore unscanned space, the most efficient solution, which is indeed observed generally, is a straight ballistic motion. Precisely, it has been observed that the turning angle after each pause is usually small\cite{r2,r10,r11,r12}. Since the direction of successive ballistic motions is strongly correlated for most of species (enjoying minimal memory skills), the small angles allows us to consider an effective 1-dimensional problem for both phases.

We now evaluate the average time needed to find a target. Following the "closed cell approach"\cite{bamberg,r18}, our problem of an infinite space with uniform target density $1/L$ can be reduced to the problem of a single target centered on a segment of size  $L$ with reflexive boundary conditions. 
Then, the instantaneous state of the searcher can be described by its position $x$ on the segment and an index $\mu$ which specifies its motion: 
\begin{itemize}
\item $\mu=\alpha$ (resp. $\beta$) corresponds to a ballistic motion of velocity $+v$ (resp. $-v$ )
\item $\mu=\gamma$ (resp. $\delta$) corresponds to a diffusive motion, switching only to a ballistic motion of velocity $ +v $ (resp. $-v$ )
\end{itemize}
The mean first passage time at the target, starting from state $(x,\mu)$ is denoted $t(x,\mu)$. Using the  Backward Chapman-Kolmogorov differential equation for the conditionnal densities\cite{r13,r14}, we obtain the following system satisfied by the $t(x,\mu)$:

\begin{equation}
\displaystyle\left\{
\begin{array}{lll}
\disp v\frac{\partial t(x,\alpha)}{\partial x}+f_2\left[t(x,\gamma)-t(x,\alpha)\right] &=& -1  \\
\disp -v\frac{\partial t(x,\beta)}{\partial x}+f_2\left[t(x,\delta)-t(x,\beta)\right] &=& -1  \\
\disp D\frac{\partial^2t(x,\gamma)}{\partial x^2}+f_1\left[t(x,\alpha)-t(x,\gamma)\right] &=& -1  \\
 \disp D\frac{\partial^2t(x,\delta)}{\partial x^2}+f_1\left[t(x,\beta)-t(x,\delta)\right] &=& -1 

\end{array}\right.
\label{sys}
\end{equation}

Henceforth, we will consider the average search time $\langle t \rangle$ defined as the average of $t(x,\mu)$  over the initial position $x$ of the searcher, which is assumed to be uniformly distributed over the segment $[-L,L]$, and over the nature of the initial motion, equally distributed over $\alpha$ and $\beta$ (ballistic motions with velocities $\pm v$).

In the low density limit defined by $
L\gg\frac{v}{f_2},\sqrt{\frac{D}{f_1}},\sqrt{\frac{f_2D}{f_1v}}$ the system (\ref{sys}) leads, after some calculation, to:
\begin{equation}
\langle t \rangle=\frac{L}{2\sqrt{D}}\left(\frac{1}{f_1}+\frac{1}{f_2}\right)\frac{\tau f_2^2+2f_1}{\sqrt{\tau f_2^2+4f_1}}
\end{equation}
where $\tau=D/v^2$ and $1/L$ is the target density. The linear dependence on $L$ (the typical inter-target distance) ensures that this combined strategy is much more efficient than a purely diffusive strategy\cite{r15} which would scale like $L^2$. The average search time $\langle t \rangle$ presents a single minimum with respect to $f_1$ and $f_2$, defined by the following equations:

\be\label{eq3}
\left\{\begin{array}{ll}
\disp f_1=f_{1{\rm max}}\\
\disp f_2^5+\frac{6}{\tau}f_2^3f_1-\frac{8}{\tau^2}f_1^3=0
\end{array}\right.
\end{equation}

This minimum takes a simple form in two different regimes which lead to similar asymptotic:

First, if $f_{1{\rm max}}\ll1/\tau$, the optimal frequencies are such that $f_1=f_{1{\rm max}}$ and:
\be\label{eq4}
f_2=\left(\frac{4}{3\tau}\right)^{1/3}f_1^{2/3}
\end{equation}
In this regime, denoted ${\cal S }$ (for search), one has $f_1<f_2$: the predator spends more time searching than moving.

Second, if $f_{1{\rm max}}\gg1/\tau$ the optimal frequencies are such that $f_1=f_{1{\rm max}}$  and :
\be\label{eq5}
f_2=\left(\frac{2\sqrt{2}}{\tau}\right)^{1/3}f_1^{3/5}
\end{equation}
In this regime, denoted by ${\cal M}$ (for move), one has $f_1>f_2$ and the predator spends more time moving. Note that the exponents 2/3 and 3/5 are numerically very close, and we do not expect to distinguish them experimentally. 

We now briefly  comment on the structure of  optimal trajectories. The threshold value $1/\tau$ has the meaning of an overlap limit: $\tau$  is the duration for which the typical distance covered is the same in both states 1 and 2. In order to study the connexity of these trajectories, we introduce the ratio $R$ of the length scanned in phase 1 over the distance covered during a phase 2 move $R=\sqrt{D/f_1}\times f_2/v$.   As for the  ${\cal S}$, one has $ R\approx(f_1\tau)^{1/6}<1$ so that there is no overlap. It is noteworthy that in this regime, $R$ can be small, which means that the overall scanned space is not connex but displays holes of unvisited space. Conversely, in regime ${\cal M}$, one has $ R\approx(f_1\tau)^{1/10}>1$ and overlap may occur; nevertheless the power 1/10 ensures that $R$ is never much larger than 1 and the optimal trajectory always explores unscanned spaces of significant size.

We now compare our model with experimental data extracted from O'Brien et al.\cite{r2} and Kramer et al.\cite{r3} which provide the average duration (and therefore its inverse, the rate) of search and motion phases for 18 different species (as various as planktivorous fish\cite{r10}, ground foraging birds\cite{r11,r16} or lizards\cite{r17}) performing a saltatory search behaviour. The corresponding rates $f_1$ range from 0.1 to 100 Hz, with no systematic correlation with the animal size. Note that the prefactor involving $\tau$  in equation (\ref{eq4}) and (\ref{eq5}) depends a priori on the species and it seems difficult to evaluate it directly from observations, as $D$ was defined in a phenomenological manner. Nevertheless, as we proceed to show, this characteristic time roughly assumes only two values. 

\begin{figure}
\scalebox{0.22}{
\includegraphics{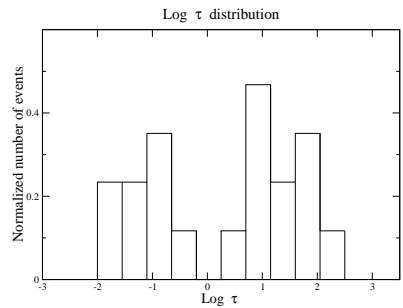}}
\caption{Log$(\tau)$ distribution for saltatory search behaviours. The first peak (around t=0.1s.) corresponds to foragers in the regime {\cal S}, the second one (around t=25s.) corresponds to foragers in regime {\cal M}.}
\label{fig2}
\end{figure}

Using the exact relation (\ref{eq3}) between $f_1$, $f_2$ and $\tau$  at the minimum, we compute the value of $\tau$ for each species. The distribution of ${\rm Log}(\tau)$, given in figure \ref{fig2}, is quite unexpectedly bimodal. Remarkably, we observed that the first peak (around t=0.1s.) corresponds to foragers in the regime {\cal S} and that  the second one (around t=25s.) corresponds to foragers in regime {\cal M}. Therefore the caracteristic time  $\tau$  defined in our model appears as a tool characterizing distinctly two subclasses of foragers: a set S of animals in the regime ${\cal S}$, and a set M in the regime ${\cal M}$. Since fluctuations of  $\tau$ are  small within each set (the fluctuations of ${\rm Log}(\tau)$ are of order 1), we can perform a comparative analysis of these data within each set. 
\begin{figure}
\scalebox{0.35}{
\includegraphics{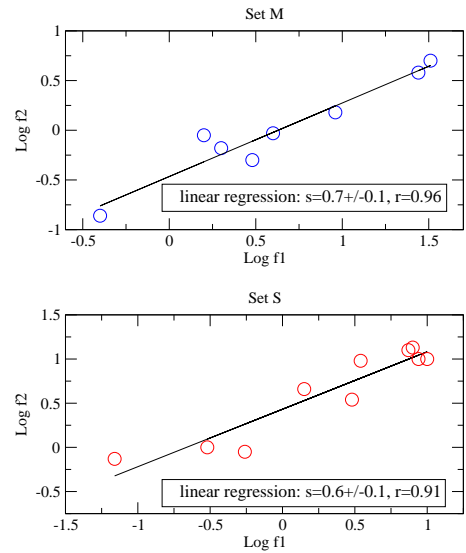}}
\caption{Log-Log plot of experimental data\cite{r2,r3} of saltatory search behaviours, and their linear regression.  }
\label{fig3}
\end{figure}

The Log-Log plots of the  $f_1$ and $f_2$   data of sets S and M are shown in figure \ref{fig3}. Their linear regression shows that both sets are strongly correlated (with a coefficient $r>0.9$) and that their slopes are in agreement with our theoretical prediction, namely 3/5 for M (experimental slope: $0.7\pm 0.1$) and 2/3 for S (experimental slope: $0.6\pm0.1$), taking account of the uncontrolled accuracy of experimental measurements. These results suggest that a wide variety of species indeed minimize their search time for preys according to a strategy described by our model, indicating that our diffusive modelling of the search phase is appropriate. This analysis also puts forward the hypothesis that the kinetics of the trajectories is a prevailing factor which is optimized by natural selection. 

\begin{figure}
\scalebox{0.22}{
\includegraphics{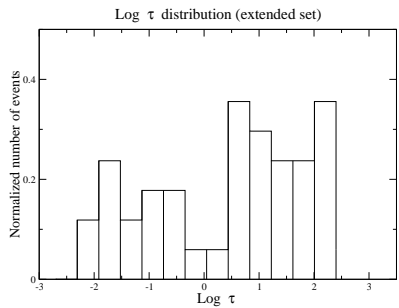}}
\caption{Log$(\tau)$ distribution for non specific saltatory behaviours.}
\label{fig4}
\end{figure}

\begin{figure}
\scalebox{0.35}{
\includegraphics{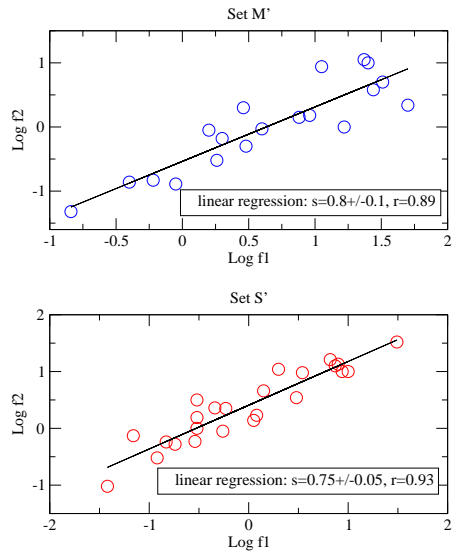}}
\caption{Log-Log plot of experimental data\cite{r2,r3} of non specific saltatory behaviours and their linear regression.}
\label{fig5}
\end{figure}

Furthermore, using the extensive data of O'Brien et al.\cite{r2} and Kramer et al\cite{r3}, we have extended this analysis to the case of animals performing a saltatory behaviour observed in non specific search behaviours, such as fleeing, food carrying, or simple displacement. The distribution of ${\rm Log}(\tau)$ over these species is still bimodal (see figure \ref{fig4})  and highly similar to the case of figure 2, and leads to the definition of extended sets S' and M' as before. The corresponding Log-Log plots, presented in figure \ref{fig5}, show that the correlations between frequencies are maintained, and that the agreement with equations (\ref{eq4}) and (\ref{eq5}) is still significant: we find an experimental slope of 0.8 for both sets S' and M'. This suggests that all displacements of these species are conditioned by a search-like mechanism which could be either the search for a prey, or for a potential threat, as for example a hint of a nearby predator. Such an attitude could also be interpreted as a "behavioural economy" like argument, which would allow  the animals  to minimize the number of behaviours to learn.

In summary, in this letter we have raised the question of determining the fastest strategy for finding a hidden target. This question of general scope has been tackled through the example of animals searching for food, for which numerous experimental data are available.  We have proposed and solved analytically a two states stochastic model of target search, based on a diffusive modeling of the perception mechanism. This model provides a power law relationship  between the characteristic times spent in each state, which satisfactorily fits experimental foraging data. 
Our findings suggest that in various behavioural contexts, saltatory animals adopt the intermittent motion which optimizes the search time of randomly hidden targets. We believe that this kind of search behaviour modeling could be extended to other situations, including human behaviours.

Thanks to  J-M Victor for fruitful discussions and to M. Barbi, R. Chitra, V. Fourcassi{\' e}, M. Jardat,  J-F Joanny , A. Lemarchand, E. Maisonhaute and P. Viot for critical reading of this manuscript.

\end{document}